\documentclass[12pt]{article}

\usepackage[x11names]{xcolor}
\usepackage{tikz}
\usepackage{cite}
\usepackage{comment}
\usepackage{hyperref}
\usetikzlibrary{decorations.pathreplacing,decorations.markings}
\usepgflibrary{arrows}
\usepgflibrary[arrows]
\usetikzlibrary{arrows}
\usetikzlibrary{positioning}
\usetikzlibrary{cd}
\usetikzlibrary{external} 
\usepackage{graphicx}
\usepackage{amsmath}
\usepackage{amssymb}
\usepackage{subsupscripts}
\usepackage{braket}
\usepackage{amsthm}
\usepackage{xcolor}
\usepackage{subcaption}

\makeatletter

\@addtoreset{equation}{section}
\makeatother
\newcommand{\ba}{\begin{eqnarray}}
\newcommand{\ea}{\end{eqnarray}}

\usetikzlibrary{decorations.pathreplacing,decorations.markings}
\tikzset{
  on each segment/.style={
    decorate,
    decoration={
      show path construction,
      moveto code={},
      lineto code={
        \path [#1]
        (\tikzinputsegmentfirst) -- (\tikzinputsegmentlast);
      },
      curveto code={
        \path [#1] (\tikzinputsegmentfirst)
        .. controls
        (\tikzinputsegmentsupporta) and (\tikzinputsegmentsupportb)
        ..
        (\tikzinputsegmentlast);
      },
      closepath code={
        \path [#1]
        (\tikzinputsegmentfirst) -- (\tikzinputsegmentlast);
      },
    },
  },
  mid arrow/.style={postaction={decorate,decoration={
        markings,
        mark=at position .5 with {\arrow[#1]{stealth}}
      }}},
}

\tikzstyle{decision} = [diamond, draw, fill=blue!20, 
    text width=4.5em, text badly centered, node distance=3cm, inner sep=0pt]
\tikzstyle{block} = [rectangle, draw, fill=blue!20, 
    text width=8em, text centered, rounded corners, minimum height=2em]
\tikzstyle{line} = [draw, -latex']
\tikzstyle{cloud} = [draw, ellipse,fill=red!20, node distance=3cm,
    minimum height=2em, text centered]

\newcommand{\dket}[1]{\ket{#1}\!\rangle}
\newcommand{\dbra}[1]{\langle\!\bra{#1}}

 \def\d{\delta}

 \def\p{\partial}
 
 \def\a{\alpha}
 \def\b{\beta}
 
 \def\d{\delta}
 \def\e{\varepsilon}
 
 \def\th{\theta}
 
 \def\k{\kappa}
 \def\l{\lambda}
 
 \def\n{\nu}

 \def\s{\sigma}
 \def\t{\tau}
 \def\th{\theta}
 \def\z{\zeta }

 \def\D{\Delta}

\def\CF{{\mathcal{F}}}

\def\CL{{\mathcal{L}}}
\def\CM{{\mathcal{M}}}
\def\CN{{\mathcal{N}}}

\def\CS{{\mathcal{S}}}

\def\CU{{\mathcal{U}}}

\def\CY{{\mathcal{Y}}}

\def\la{\left\langle}
\def\ra{\right\rangle}

\def\implies{\quad\Rightarrow\quad}

\def\vphi{\varphi}

\def\brho{\bar\rho}

\def\CS{\mathcal{S}}

\def\qf{\mathfrak{q}}

\def\Zv{Z^{(\text{vect.})}}

\def\Zf{Z^{(\text{f})}}
\def\Zaf{Z^{(\text{af})}}

\def\ZCS{Z^{(\text{CS})}}

\def\CZinst{\mathcal{Z}_I}

\def\vac{\emptyset}
\def\res{\mathop{\text{Res}}}

\def\bk{{\boldsymbol{k}}}
\def\bnu{{\boldsymbol{\nu}}}

\def\mZ{\mathbb{Z}}
\def\mR{\mathbb{R}}
\def\mC{\mathbb{C}}

\def\bN{\bar{N}}

\def\gl{\mathfrak{gl}}
\def\sl{\mathfrak{sl}}

\def\Uqsl{U_q(\widehat{\mathfrak{sl}(2)})}

\def\Abox{{\tikz[scale=0.007cm] \draw (0,0) rectangle (1,1);}}
\def\sAbox{{\tikz[scale=0.005cm] \draw (0,0) rectangle (1,1);}}

\def\bmu{{\boldsymbol{\mu}}}

\def\fund{^{(\text{f})}}
\def\afund{^{(\text{af})}}

\begin{document}
\begin{titlepage}
\vspace*{-2cm}
\begin{flushright}
\end{flushright}
\vspace*{1cm}

\begin{center}
    {\Huge  Engineering 3D $\CN=2$ theories using the\\
    \vskip .5cm
    quantum affine $\sl(2)$ algebra}
    \vskip 2cm
    {\Large Jean-Emile Bourgine}\\
    \vskip 1cm
    {\it ARC Centre of Excellence for Mathematical and Statistical Frontiers (ACEMS)}\\
    {\it School of Mathematics and Statistics}\\
    {\it University of Melbourne}\\
    {\it Parkville, Victoria 3010, Australia}\\
    \vskip 1cm    
    {\it Center for Quantum Spacetime (CQUeST)}\\
    {\it Sogang University}\\
    {\it Seoul, 121-742, South Korea}
    \vskip 1cm
    \texttt{bourgine@kias.re.kr}
\end{center}
\vfill
\begin{abstract}
The algebraic engineering technique is applied to a class of 3D $\CN=2$ gauge theories on the omega-deformed background $\mR_\e^2\times S^1$. The vortex partition function and the fundamental qq-character are obtained from a network of intertwiners between representations of the shifted (or asymptotic) quantum affine $\sl(2)$ algebra. This network involves two types of representations, the prefundamental representation of Hernandez-Jimbo, and a new vertex representation acting on a bosonic Fock space. The brane system associated to this network is identified: D3 branes carry the prefundamental module while NS5-branes (+D5) support the Fock module. In the process, we highlight the role of shifted quantum algebras in implementing the Higgsing procedure.
\end{abstract}
\vfill
\end{titlepage}

\setcounter{footnote}{0}

\newpage


\section{Introduction}
In the \textit{geometric engineering}, supersymmetric gauge theories are typically obtained by compactification of string, M-, or F- theory on a Calabi-Yau geometry \cite{Katz1996}. In contrast, the \textit{algebraic engineering} reconstructs the expression of their localized BPS-observables using the representation theory of a quantum group \cite{Mironov2016,Bourgine2017b,Bourgine2017a}. These two different constructions share a common structure, namely the brane system whose low energy dynamics realize the gauge theory. This brane system encodes the toric geometry of the Calabi-Yau \cite{Leung1998}, and provides the guidelines to build a network of representations on which act the operators of the algebra.\footnote{This network is also called ``network matrix model'' in \cite{Mironov2016,Awata2016a} but we will not discuss the relation with matrix models here.} The complementarity of these two approaches explain their success, in particular for the description of the Bethe/gauge and the BPS/CFT correspondences.


The algebraic engineering has been introduced for the first time in the context of 5D $\CN=1$ gauge theories on the omega-deformed spacetime $\mR_{\e_1}^2\times\mR_{\e_2}^2\times S^1$ \cite{Mironov2016,Awata2016a,Bourgine2017b}.\footnote{These theories reduce to 4D $\CN=2$ theories in the limit $R\to0$ of the compactification radius.} The partition function of these theories are one-loop exact, but receive a tower of non-perturbative (instanton) corrections. As an alternative to the localization \cite{Nekrasov2002}, these partition functions can also be computed using a correspondence with topological strings amplitudes. And, in fact, the key ingredient of the algebraic construction is an intertwining operator of the quantum toroidal $\gl(1)$ algebra whose matrix elements coincide with the refined topological vertex  \cite{AFS}.\footnote{The quantum toroidal $\gl(1)$ algebra is known under many names, among which ``Ding-Iohara-Miki'' (or DIM) algebra \cite{Ding1997,Miki2007} has acquired some popularity in the physics literature. We settled here on the name ``quantum toroidal $\gl(1)$'' in an attempt of standardization since its higher rank versions, i.e. the quantum toroidal $\gl(n)$ algebras, enter in similar constructions pertaining to gauge theories on ALE spaces \cite{Awata2017,Bourgine:2019phm}.} The corresponding brane system is a web of $(p,q)$ 5-branes \cite{Aharony1997,Aharony1997a}, each brane being associated to a Fock representation with levels $(q,p)$ of the toroidal algebra. This algebraic construction was later refined to include other observables (Wilson loops) and extended to a large class of gauge theories in various dimensions \cite{BFMZ,Zhu2017,Foda2018,Awata2017,Zenkevich2018,Zenkevich:2019ayk,Zenkevich2020,Ghoneim2020,Bourgine2018,Bourgine2018a,Bourgine:2019phm}, thereby effectively implementing new ``topological vertex techniques'' for the corresponding brane systems. All these constructions involve toroidal (or double affine) algebras. In this letter, we present a construction based on a simpler algebra, namely the quantum affine $\sl(2)$ algebra, for a class of $\CN=2$ 3D gauge theories on $\mR_{\e}^2\times S^1$.\footnote{This technique can be trivially extended to the 2D $\CN=(2,2)$ theories obtained in the limit $R\to0$ using the Yangian double of $\sl(2)$ \cite{Khoroshkin1994,Khoroshkin1996} (just like 4D $\CN=2$ theories were addressed using the affine Yangian double of $\gl(1)$ in \cite{Bourgine2018}).}

The 3D $\CN=2$ theories considered here are obtained as a limit of 3D $\CN=2^\ast$ theories, i.e. massive deformations of 3D $\CN=4$ theories, by sending to infinity the mass of the adjoint chiral multiplet in order to decouple its fields from the theory. It has been noticed in \cite{Aganagic2013,Nedelin2017,Aprile2018} that the vortex partition function of these 3D $\CN=2^\ast$ theories can be obtained from the instanton partition functions of 5D $\CN=1$ theories through a Higgsing procedure, and that the algebraic engineering of the latter can be used to produce the observables of the former after tuning the mass parameters. It was later proposed by Zenkevich in \cite{Zenkevich2018} to use the vector representation of the toroidal algebra to obtain directly the vortex partition functions. We will comment more later on the connection between these two derivations of 3D $\CN=2^\ast$ partition functions, both based on the quantum toroidal $\gl(1)$ algebra, and the approach proposed here for 3D $\CN=2$ theories using the quantum affine $\sl(2)$ algebra. But, in a nutshell, our approach is simpler as it avoids the tuning of parameters in \cite{Nedelin2017,Aprile2018}, bypasses the Jackson q-integration performed in \cite{Zenkevich2018}, and produces no spurious factors. It is also slightly more general as it is possible to introduce any number of fundamental/antifundamental chiral matter multiplets with arbitrary masses. On the other hand, we find it necessary to employ a shifted \cite{Hernandez2020} (or asymptotic \cite{Hernandez2012}) version of the quantum affine $\sl(2)$ algebra. These shifts naturally arise in the presence of fundamental/antifundamental matter multiplets, and such fields are required in order to obtain a non-trivial vortex partition function. We will come back to the role of the shifted algebras in the companion paper \cite{Bourgine2021c} presenting the details of our construction and its relation with the algebraic engineering of 5D $\CN=1$ theories.
%

The vortex partition functions are relatively simple objects, but they have a central role in many discussions. They have been referred as ``3D holomorphic blocks'' in \cite{Beem2012} as they enter in the construction of observables for gauge theories with more complicated 3D space-times. They can also be interpreted as K-theoretic partition functions of 2D $\CN=(2,2)$ gauge theories \cite{Dimofte2010}, and correspond to the ``J-function'' of Gromov-Witten theory \cite{Givental1993,Givental2001}. The algebraic technique presented here have deep connections with the quantum chiral ring of 3D $\CN=4$ Coulomb branches \cite{Bullimore2015,Braverman2016}, and the correspondence with integrable systems described in \cite{Gadde2013}.

To be specific, our construction relies on two distinct representations that we call \textit{vertical} and \textit{horizontal} by analogy with the 5D/toroidal case \cite{AFS}. The vertical representation is an infinite dimensional highest weight representation called \textit{prefundamental} representation \cite{Hernandez2012}. It coincides with the (double and K-theoretic) COhomological Hall Algebra of the quiver variety associated to the 3D $\CN=2$ gauge theories in \cite{Fujitsuka2013} (namely a handsaw quiver with a single gauged node). Instead, the horizontal representation is a vertex representation, acting on a free boson Fock space, that appears to be new. This representation is a crucial ingredient for the derivation of intertwining operators, and thus the whole formulation of the algebraic engineering.

This paper is organized as follows. In the second section, we recall the expression of the vortex partition functions, and the associated string theory description. The third section presents our main results, namely the engineering of partition functions and qq-characters. Finally, we briefly comment on the relation with the previously known formalisms in section four.

\section{3D $\CN=2$ gauge theories}
\subsection{Vortex partition functions}\label{sec_COHA}
We consider a class of gauge theories with $\CN=2$ supersymmetry (four supercharges) defined on the omega-deformed background $\mR^2_\e\times S^1$. The omega-background parameter $\e$ combined with the radius $R$ of $S^1$ defines the quantum group parameter $q^2=e^{R\e}$ (we assume $|q|<1$). The partition function receives factorized contributions from classical, one-loop and non-perturbative vortex corrections, and we focus only on the vortices' contribution that we denote $Z_V$. It has been evaluated by Higgs branch localization in \cite{Yoshida2011,Fujitsuka2013} (see also \cite{Chen2013}).

Specifically, we focus here on $U(N)$ gauge theories with a single $\CN=2$ vector multiplet, $N\fund\geq N$ chiral multiplets of masses  $m_a\fund$ ($a=1,\cdots,N\fund$) in the fundamental representation, and $N\afund$ chiral multiplets with mass $m\afund_a$ in the antifundamental representation. Localized configurations are labeled by a set of integers $(\ell_1,\cdots,\ell_N)$ in $(1,\cdots,N_f)$ that correspond to a choice of boundary conditions for the fields. These integers label the massive vacua of the supersymmetric vortex quantum mechanics describing the theory after localization on $\mR_\e^2$. For simplicity, following \cite{Fujitsuka2013}, we take $\ell_i=i$ and denote the exponentiated fundamental masses $\nu_a\fund=e^{Rm_a\fund}$ as $\nu_i=\nu_i\fund$ for $i=1\cdots N$ (other choices correspond to $\nu_i=\nu_{\ell_i}\fund$), they form the vectors $\bnu=(\n_1,\cdots\nu_N)$ and $\bnu\fund=(\nu_{N+1}\fund,\cdots \nu_{N_f}\fund)$. We denote the exponentiated masses of the antifundamental fields $\bnu\afund=(e^{Rm\afund_1},\cdots,e^{Rm\afund_{N\afund}})$. Finally, the action is further deformed by a Chern-Simons term of level $\k$.

The vortex partition function is a sum over the abelianized vortex numbers $\bk=(k_1,\cdots,k_N)\in(\mZ^{\geq0})^N$, $k_i$ being the vorticity of each subgroup $U(1)\subset U(N)$ on the $i$th diagonal, \footnote{The proper vortex partition function differs from the equivariant partition function presented here by extra harmless factors that can be found in \cite{Fujitsuka2013}. These factors can be absorbed in a redefinition of the FI parameters and a shift of the Chern-Simons level, they are vanishing in the 2D limit $R\to0$.}
\begin{equation}\label{CZ_V}
Z_V[U(N),N_f,\bN_f]=\sum_{\bk}\prod_{i=1}^N\qf_i^{k_i}\Zv(\bk,\bnu)\ZCS(\bk,\bnu,\k)\Zf(\bk,\bnu,\bnu\fund)\Zaf(\bk,\bnu,\bnu\afund).
\end{equation} 
Each configuration is weighted by the complexified FI-parameters $\qf_i$, and the summands factorize into contributions from the various $\CN=2$ multiplets and the Chern-Simons term,\footnote{The notation $\Zv(\bk,\bnu)$ is a little abusive here as it contains also the contribution from the first $N$ fundamental chiral multiplets with masses $\bnu$.}
\begin{align}\label{vortex_UN}
\begin{split}
&\Zv(\bk,\bnu)=\prod_{i,j=1}^NN_{k_i,k_j}(\nu_i/\nu_j)^{-1},\quad \ZCS(\bk,\bnu,\k)=\prod_{i=1}^N (\nu_iq^{k_i(k_i-1)})^{\k},\\
&\Zf(\bk,\bnu,\bnu\fund)=\prod_{i=1}^N\prod_{a=N+1}^{N\fund}(\nu_i/\nu\fund_a;q^2)_{k_i}^{-1},\quad \Zaf(\bk,\bnu,\bnu\afund)=\prod_{i=1}^N\prod_{a=1}^{N\afund}(\nu_i/\nu\afund_a;q^2)_{k_i},
\end{split}
\end{align}
where we have introduced a 3D version of the Nekrasov factor expressed using the Pochhammer symbol,\footnote{Note the important reflection property
\begin{equation}\label{prop_Nkk}
\dfrac{N_{k,k'}(\a)}{N_{0,k'}(\a)}=q^{2kk'}\dfrac{N_{0,k}(q^{-2}\a^{-1})}{N_{k',k}(q^{-2}\a^{-1})}.
\end{equation}}
\begin{equation}\label{def_Nkk}
N_{k,k'}(\a)=(\a q^{2k-2k'+2};q^2)_{k'},\quad (z,q^2)_k=\prod_{j=0}^{k-1}(1-zq^{2j}).
\end{equation} 
Remarkably, the vortex partition function does not depend on the gauge coupling of the theory.

\subsection{Higgsing 5D $\CN=1$ theories and branes systems}\label{sec_string}
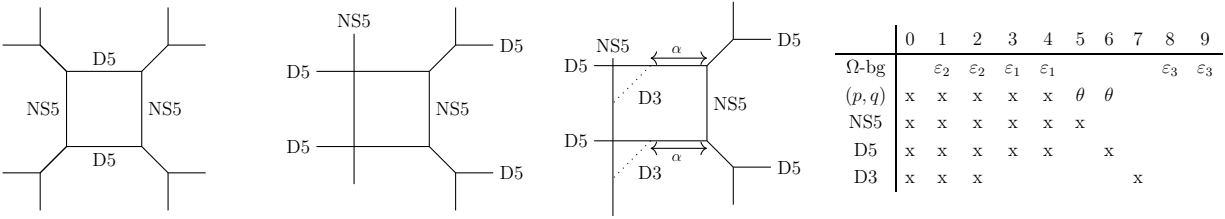
\begin{figure}
\begin{subfigure}[c]{0.2\textwidth}
\begin{tikzpicture}[scale=.5]
\draw (-1.7,-.7) -- (-.7,-.7) -- (0,0) -- (2,0) -- (2.7,-.7) -- (3.7,-.7);
\draw (-.7,-1.7) -- (-.7,-.7) -- (0,0) -- (0,2) -- (-.7,2.7) -- (-1.7,2.7);
\draw (-.7,3.7) -- (-.7,2.7) -- (0,2) -- (2,2) -- (2.7,2.7) -- (2.7,3.7);
\draw (2,0) -- (2,2);
\draw (2.7,2.7) -- (3.7,2.7);
\draw (2.7,-.7) -- (2.7,-1.7);
\node[scale=.6,left] at (0,1) {NS5};
\node[scale=.6,below] at (1,0) {D5};
\node[scale=.6,right] at (2,1) {NS5};
\node[scale=.6,above] at (1,2) {D5};
\end{tikzpicture}
\end{subfigure}
\begin{subfigure}{0.2\textwidth}
\begin{tikzpicture}[scale=.5]
\draw (-1,0) -- (2,0) -- (2.7,-.7) -- (3.7,-.7);
\draw (-1,2) -- (2,2) -- (2.7,2.7) -- (2.7,3.7);
\draw (2,0) -- (2,2);
\draw (0,-1) -- (0,3);
\draw (2.7,2.7) -- (3.7,2.7);
\draw (2.7,-.7) -- (2.7,-1.7);
\node[scale=.6,above] at (0,3) {NS5};
\node[scale=.6,left] at (-1,2) {D5};
\node[scale=.6,left] at (-1,0) {D5};
\node[scale=.6,right] at (2,1) {NS5};
\node[scale=.6,right] at (3.7,2.7) {D5};
\node[scale=.6,right] at (3.7,-.7) {D5};
\end{tikzpicture}
\end{subfigure}
\begin{subfigure}{0.2\textwidth}
\begin{tikzpicture}[scale=.5]
\draw (-1,0) -- (2,0) -- (2.7,-.7) -- (3.7,-.7);
\draw (-1,2) -- (2,2) -- (2.7,2.7) -- (2.7,3.7);
\draw (2,0) -- (2,2);
\draw (-.5,-2) -- (-.5,2.2);
\draw (2.7,2.7) -- (3.7,2.7);
\draw (2.7,-.7) -- (2.7,-1.7);
\draw[dotted] (0.5,0) -- (-.5,-1);
\draw[dotted] (0.5,2) -- (-.5,1);
\draw[<->] (0.5,2.2) -- (2,2.2);
\draw[<->] (0.5,-.2) -- (2,-.2);
\node[scale=.5,above] at (1.2,2.2) {$\a$};
\node[scale=.5,below] at (1.2,-.2) {$\a$};
\node[scale=.6,above] at (-.5,2.2) {NS5};
\node[scale=.6,left] at (-1,2) {D5};
\node[scale=.6,left] at (-1,0) {D5};
\node[scale=.6,below right] at (0,-.5) {D3};
\node[scale=.6,below right] at (0,1.5) {D3};
\node[scale=.6,right] at (2,1) {NS5};
\node[scale=.6,right] at (3.7,2.7) {D5};
\node[scale=.6,right] at (3.7,-.7) {D5};
\end{tikzpicture}
\end{subfigure}
\begin{subfigure}{0.2\textwidth}
\scalebox{.6}{\begin{tabular}[b]{c|cccccccccc}
& 0 & 1 & 2 & 3 & 4 & 5 & 6 & 7 & 8 & 9\\
\hline
$\Omega$-bg & & $\e_2$ & $\e_2$ & $\e_1$ & $\e_1$ & & & & $\e_3$ & $\e_3$\\
$(p,q)$ & x & x & x & x & x & $\th$ & $\th$ & & & \\
NS5 & x & x & x & x & x & x & & & & \\
D5 & x & x & x & x & x & & x & & & \\ 
D3 & x & x & x & & & & & x & &\\
\end{tabular}}
\end{subfigure}
\caption{Higgsing procedure on the $(p,q)$-brane web describing the 5D $\CN=1$ U(2) gauge theory with four flavors.}
\label{figHiggs}
\end{figure}

To describe the Higgsing procedure, we start from a 5D $\CN=1$ $U(N)$ gauge theory, defined in the omega-background $\mR_{\e_1}^2\times\mR_{\e_2}^2\times S_R^1$, with $N$ hypermultiplets and $N\afund$ antifundamental hypermultiplets. We denote the exponentiated masses of the fields $\mu\fund_i$ and $\mu\afund_a$ respectively, with $i=1\cdots N$ and $a=1\cdots N\afund$. The theory is considered on the Coulomb branch where the gauge group $U(N)$ is broken to $U(1)^N$ by the vacuum expectation values (vev) of the vector multiplet's adjoint scalar. The exp. Coulomb branch parameters are denoted $v_i$. Ignoring perturbative and 1-loop factors, we focus on the instantons contributions to the partition function $\CZinst[U(N),N,\bN_f]$ that have been computed exactly by localization in \cite{Nekrasov2002}. In addition to $v_i$, $\mu\fund_i$ and $\mu\afund_a$, the result also involves the parameters  $(q_1,q_2)=(e^{R\e_1},e^{R\e_2})$, and takes the form of an expansion in powers of the (renormalized) exp. gauge coupling $\qf_I$.

The Higgs branch and the Coulomb branch of the 5D theory meet at the point $\mu_i\fund=v_i$. The non-abelian vortices on the Higgs branch have worldvolume $\mR_{\e_2}\times S_R^1$, they are described by a 3D $\CN=2^\ast$ theory that will later reduce to our 3D $\CN=2$ theory in the limit of infinite $\e_1$-parameter. Due to the gauge/vortex correspondence, the 3D vortex partition function coincides with the 5D instanton partition function after tuning the mass parameters to the point $q_1^{-1}q_2^{-1}\mu_i\fund=q_1v_i$ \cite{Aganagic2013,Nedelin2017,Aprile2018}. The summation over instanton configurations is truncated to $N$-tuples Young diagrams with only a single column. The height of this column is identified with the vortex number $k_i$. The summands greatly simplify, and can be written using the 3D building blocks given in \ref{vortex_UN},
\begin{equation}\label{CZ_Higgsed}
\CZinst[U(N),N,\bN_f|\k]=\sum_{\bk}\qf_V^{|\bk|}\ZCS(\bk,\bnu,\k')\Zv(\bk,\bnu)\Zaf(\bk,\bnu,\bmu\afund)\times\prod_{i,i'=1}^N\dfrac{N_{k_i,k_{i'}}(q_1\nu_i/\nu_{i'})}{(q_1q_2\nu_{i'}/\nu_i;q_2)_{k_{i'}}},
\end{equation} 
where we naturally identified the omega-background parameters $q_2=q^2$, and the Coulomb branch vevs $v_i$ with the exp. mass $\nu_i$ of the fundamental chiral multiplets. In addition, the instanton counting parameter has been rescaled into $\qf_V=\qf_Iq_1^{-N}\prod_a(-\mu_a\afund)/\prod_i(-\nu_i)$, and the Chern-Simons level received the correction $\k'=\k+N-N\afund$. The field content of the 3D theory can be deduced from the expression \ref{CZ_Higgsed}. It consists in a single $U(N)$ vector multiplet and $N$ fundamental chiral multiplets of mass $\nu_i$ (contributing to $\Zv(\bk,\bnu)$), $N\afund$ antifundamental chiral multiplets of mass $\mu_a\afund$ (contributing to $\Zaf(\bk,\bnu,\bmu\afund)$), and one $U(N)$ adjoint chiral multiplet of exp. mass $\mu=q_1^{-1}$.\footnote{The identification of the contribution for the adjoint chiral multiplet relies on the observation made in \cite{Fujitsuka2013} that the 3D $\CN=2^\ast$ partition function with exp. adjoint mass $q_1^{-1}$ reads
\begin{equation}
\sum_{\bk}\qf_V^{|\bk|}\prod_{i,i'}\dfrac{N_{k_i,k_{i'}}(q_1\nu_i/\nu_{i'})}{N_{k_i,k_{i'}}(\nu_i/\nu_{i'})}.
\end{equation} 
This theory contains a $U(N)$ vector multiplet, a chiral adjoint multiplet of exp. mass $\mu$ (both descending from the $\CN=4$ vector multiplet), and $N$ fundamental chiral multiplets of exp. mass $\nu_i$ and $N$ antifundamental chiral multiplets of exp. mass $q_1^{-1}\nu_i$ (from the $\CN=4$ hypermultiplet). Supposedly, the role of the factors $(q_1q_2\nu_{i'}/\nu_i;q_2)_{k_{i'}}$ in the denominator of \ref{CZ_Higgsed} is to remove the contribution of the extra $N$ antifundamental chiral multiplets. However, the presence of the shift $q_2$ in the argument (that brings the extra factors $(1-q_1q_2^{2k_{i'}}\nu_{i'}/\nu_i)/(1-q_1\nu_{i'}/\nu_i)$) is not clear. In any case, it will play no role here since it disappears in the limit $q_1\to0$.}

To recover our 3D $\CN=2$ theories, the adjoint chiral multiplet is decoupled by sending its mass $m=-\e_1\to\infty$. This procedure corresponds to take the limit $q_1\to0$ of the Higgsed instanton partition function with $q_2=q^2$ and $\qf_V$ fixed. Then, the  instanton partition function reduces to the vortex partition function $Z_V[U(N),N,N\afund|\k']$ given in \ref{CZ_V}, with the exp. FI parameter $\qf_V$, and the antifundamental masses $\bnu\afund=\bmu\afund$ inherited from the 5D theory.\footnote{Similar results are obtained in the limit $q_1\to\infty$, but with the opposite shift $\k\to\k-N-N\afund$ of the Chern-Simons level. Higgsing instead the 5D $\CN=1$ antifundamental hypermultiplets by setting $N\afund=N$ and $\mu_i\afund=v_iq_1$, the shifts of the Chern-Simons level is no longer necessary, the FI parameter is simply obtained as $\qf_V=q_1^{-N}\qf_I$ and the fundamental chiral masses are $\nu\fund_a=\mu_a\fund q_1^{-1}q_2^{-1}$. These shifts do not appear to have any physical significance, but instead relate to our choice of K-theoretic vortex and instanton partition functions.}

The description of the Higgsing procedure in terms of brane systems can be found in \cite{Nedelin2017,Aprile2018} (it is also closely related to the construction of surface defects in 5D and 4D gauge theories \cite{Hanany1997,Gadde2013,Alday2009}). To simplify the picture (both literally and figuratively), we restrict ourselves to the case of a U(2) gauge theory with four hypermultiplets, $N=2$ in the fundamental and $N\afund=2$ in the antifundamental, and no Chern-Simons term. The $(p,q)$-branes web realizing this particular theory is represented on figure \ref{figHiggs} (left). Tuning the masses of the fundamental hypermultiplets, the external D5-branes on the left align with the internal ones (figure \ref{figHiggs} middle), which, in turns, allows us to pull out the NS5-brane, creating in the process two D3-branes attached to the D5-branes (figure \ref{figHiggs} right). In this configuration, the $(p,q)$-branes are extended along the directions 01234, together with a segment in the (56)-plane, while the D3-branes extend in the direction 0127.

To analyze the field content of the gauge theory describing the low-energy limit of this brane system, it is easier to start from the configuration with $\a=0$ for which the D3-branes end on the NS5-brane. On these two D3 branes lives a 3D $\CN=4$ gauge theory with a $U(2)$ gauge group broken into $U(1)\times U(1)$ by the distances between them along the directions 345. These distances correspond to the vev of the three scalar fields in the vector multiplet. The gauge coupling is proportional to the inverse of the length of the D3-branes. Turning on a fluxtrap to produce the omega-background with parameters $\e_1$ in the plane 34 and $\e_2$ in the plane 12 (see the table of figure \ref{figHiggs}), the $\CN=4$ vector multiplet is broken into a $\CN=2$ vector multiplet (with a real scalar corresponding to $x^5$) and a $\CN=2$ adjoint chiral multiplet containing the two remaining scalars (directions 34). Due to the omega-deformation, the chiral multiplet acquires a twisted mass $m=-\e_1$ \cite{Hellerman2011}.

To complete the picture, it remains to take into account the presence of the semi-infinite D5-branes in the (56)-plane. Each brane provides a massive $\CN=2$ chiral multiplet, either in the fundamental or antifundamental representation (depending on whether the brane is on the left or on the right of the NS5-brane). Their masses correspond to the D5-branes' positions along the fifth direction, i.e. the fundamental multiplets have masses $\log v_i$ and the antifundamental ones $\log \mu_a\afund$. Turning back $\a>0$, the endpoints of the D3-branes move along the D5-branes, and the distance $\a$, once complexified and renormalized, provides the FI parameter of the 3D gauge theory. 


\section{Algebraic engineering}
The goal of this section is to establish the free field formulas characteristic of the algebraic engineering. In particular, the vortex partition functions discussed earlier take the form of sums of products of free field correlators,
\begin{equation}
Z_V[U(N),N\fund,N\afund]\sim\sum_{\bk}\bra{\vac}\Phi_{k_1}^\ast\cdots\Phi_{k_n}^\ast\ket{\vac}\bra{\vac}\Phi_{k_1}\cdots\Phi_{k_n}\ket{\vac}.
\end{equation} 
In the algebraic formalism, the summation over the vortex numbers $\bk$ corresponds to trace over the states in the vertical representation defined below. On the other hand, the correlators of vertex operators $\Phi_k$, $\Phi_k^\ast$ are evaluated in the Fock module of the horizontal representation. These operators are determined from an intertwining condition between the horizontal representation and the tensor product vertical $\otimes$ horizontal.

\subsection{Shifted quantum affine $\sl(2)$ algebra and representations}\label{sec_prefund}
The quantum affine algebra of $\sl(2)$ with parameter $q$ is traditionally denoted $\Uqsl$, it will be abbreviated by $\CU$ here. Like in the toroidal case, we employ the Drinfeld presentation given in terms of the currents
\begin{equation}\label{Uqsl_currents}
X^\pm(z)=\sum_{k\in\mathbb{Z}}z^{-k}X^\pm_k,\quad \Psi^\pm(z)=\sum_{k\geq0}z^{\mp k}\Psi_{\pm k}^\pm,
\end{equation} 
and the central element $q^c$. We refer to \cite{Ding1993} for the dictionary between this presentation and the RTT presentation used in the field of integrable systems. The algebraic relations obeyed by the modes $X_k^\pm$ and $\Psi_k^\pm$ are well-known, they can found e.g. in \cite{Hernandez2012}. Once summed up, they takes the compact form of exchange relations among the currents,
\begin{align}
\begin{split}\label{def_Uqsl2}
&[\Psi^\pm(z),\Psi^\pm(w)]=0,\quad \Psi^+(z)\Psi^-(w)=\dfrac{G(q^cz/w)}{G(q^{-c}z/w)}\Psi^-(w)\Psi^+(z),\\
&\Psi^+(z)X^\pm(w)=G(q^{\pm c/2}z/w)^{\pm1}X^\pm(w)\Psi^+(z),\quad \Psi^-(z)X^\pm(w)=G(q^{\mp c/2}z/w)^{\pm1}X^\pm(w)\Psi^-(z),\\
&X^\pm(z)X^\pm(w)=G(z/w)^{\pm1}X^\pm(w)X^\pm(z),\\
&[X^+(z),X^-(w)]=\dfrac1{q-q^{-1}}\left(\d(q^{-c}z/w)\Psi^+(q^{c/2}w)-\d(q^c z/w)\Psi^-(q^{-c/2}w)\right),
\end{split}
\end{align}
involving the structure function
\begin{equation}\label{def_G}
G(z)=q^2\dfrac{z-q^{-2}}{z-q^2}.
\end{equation} 
The modes of the currents $\Psi^\pm(z)$ generate the Cartan subalgebra. From \ref{def_Uqsl2}, we deduce that their zero modes satisfy the exchange relations $\Psi_0^+ X^\pm_k=q^{\pm2} X^\pm_k\Psi_0^+$, $\Psi_0^- X^\pm_k=q^{\mp2} X^\pm_k\Psi_0^-$ and $\Psi_0^+\Psi_0^-=\Psi_0^-\Psi_0^+$. In order to define properly the quantum affine $\sl(2)$ algebra, these relations must be supplemented by the constraint $\Psi_0^+\Psi_0^-=1$ that does not follow from the currents relations \ref{def_Uqsl2}. Omitting this extra condition, the relations \ref{def_Uqsl2} define a slightly different algebra, namely the \textbf{asymptotic} quantum affine $\sl(2)$ algebra of Hernandez and Jimbo \cite{Hernandez2012}. As we shall see, this algebra, which we denote $\CU^a$, has a richer representation theory.

Several coproducts can be introduced to confer to the algebra $\CU$ the structure of a Hopf algebra. Like in the toroidal case, we employ the Drinfeld coproduct that acts on the currents as
\begin{align}
\begin{split}\label{def_coprod}
&\D(X^+(z))=X^+(z)\otimes 1+\Psi^-(q^{c_{(1)}/2}z)\otimes X^+(q^{c_{(1)}}z),\\
&\D(X^-(z))=X^-(q^{c_{(2)}}z)\otimes \Psi^+(q^{c_{(2)}/2}z)+1\otimes X^-(z),\\
&\D(\Psi^\pm(z))=\Psi^\pm(q^{\pm c_{(2)}/2}z)\otimes\Psi^\pm(q^{\mp c_{(1)}/2}z),
\end{split}
\end{align}
with the notation $q^{c_{(1)}}=q^c\otimes 1$, $q^{c_{(2)}}=1\otimes q^c$, and $\D(q^c)=q^c\otimes q^c$.

\paragraph{Shifted algebra} The asymptotic algebra $\CU^a$ is closely related to another deformation of the algebra $\CU$, namely the shifted algebra $\CU^{\bmu}$ with parameters $\bmu=(\mu_+,\mu_-)\in\mZ\times\mZ$ \cite{Hernandez2020}. Shifted quantum affine algebras have been introduced by Finkelberg and Tsymbaliuk in their study of (K-theoretic) Coulomb branches of 3D $\CN=4$ supersymmetric gauge theories \cite{Finkelberg2017}. Not surprisingly, these algebras also play a fundamental role in our construction. The shifted quantum affine $\sl(2)$ algebra $\CU^{\bmu}$ is defined by the same relations \ref{def_Uqsl2} between the Drinfeld currents, but with a different (shifted) expansion for the Cartan currents\footnote{It is sufficient to require $\Psi_{\mp\mu_\pm}^\pm$ invertible, and that their product is central. Imposing further $\Psi_{-\mu_+}^+\Psi_{\mu_-}^-=1$ fixes the degree of freedom consisting in rescaling $X^+(z)$ (or $X^-(z)$) and both $\Psi^\pm(z)$. However, the price to pay is the proliferation of factors $(-\nu)^{1/2}$ for $\nu\in\mC^\times$ in formulas which we would like to avoid in this short paper.}
\begin{equation}\label{shifted_Psi}
\Psi^\pm(z)=\sum_{k=-\mu_\pm}^\infty z^{\mp k}\Psi_{\pm k}^\pm,\quad\text{with}\quad \Psi_{-\mu_+}^+\Psi_{\mu_-}^-=\Psi_{\mu_-}^-\Psi_{-\mu_+}^+\in\mC^\times.
\end{equation} 
When $(\mu_+,\mu_-)=(0,0)$ and imposing $\Psi_0^+\Psi_0^-=1$, the algebra obviously reproduces $\CU=\CU^{(0,0)}$. Moreover, when $\mu_\pm\leq0$, the shifted algebra is a special case of the asymptotic algebra $\CU^a$. For any $\bmu$, the Drinfeld coproduct \ref{def_coprod} defines a homomorphism of algebras $\CU^{\bmu+\bmu'}\to\CU^{\bmu}\otimes\CU^{\bmu'}$ (after completion) \cite{Hernandez2020}.

Let $P(z)$ be a polynomial, or a finite Laurent series, in $z$, with the asymptotic behavior $P(z)\sim_0\a_0 z^{-\mu_0}$ and $P(z)\sim_\infty \a_\infty z^{\mu_\infty}$ encoded in the 2-vector $\bmu_P=(\mu_\infty,\mu_0)$. We define two homomorphisms of algebras $\iota_P:\CU^{\bmu}\to \CU^{\bmu+\bmu_P}$ and $\iota_P^\ast:\CU^{\bmu}\to \CU^{\bmu+\bmu_P}$ as follows,
\begin{align}
\begin{split}\label{def_iota}
&\iota_P: X^+(z)\to P(z)X^+(z),\quad X^-(z)\to X^-(z),\quad \Psi^\pm(z)\to P(q^{\pm c/2}z)\Psi^\pm(z),\\
&\iota_P^\ast: X^+(z)\to X^+(z),\quad X^-(z)\to P(z)X^-(z),\quad \Psi^\pm(z)\to P(q^{\mp c/2}z)\Psi^\pm(z).
\end{split}
\end{align}
The homomorphisms $\iota_P$ and $\iota_P^\ast$ form an obvious abelian semi-group since $\iota_{P_1}\iota_{P_2}=\iota_{P_1P_2}$, $\iota_{P_1}^\ast\iota_{P_2}^\ast=\iota_{P_1P_2}^\ast$ and $\iota_{P_1}\iota_{P_2}^\ast=\iota_{P_2}^\ast\iota_{P_1}$. Following \cite{Finkelberg2017}, we call them \textit{shift homomorphisms}. They can be used to define the representation of a shifted algebra from a representation $\rho$ of the original algebra $\CU$ by multiplying the action of $\rho(X^+(z))$ and $\rho(\Psi^\pm(z))$ with $P(z)$ and $\rho(P(q^{\pm c/2}z))$ (resp. the action of $\rho(X^-(z))$ and $\rho(\Psi^\pm(z))$ with $P(z)$ and $\rho(P(q^{\mp c/2}z))$ for $\iota_P^\ast$). We denote $\iota_P\rho$ and $\iota_P^\ast\rho$ these representations of $\CU^{\bmu_P}$, and call them \textit{shifted representations}.

%
%

%

\paragraph{Vertical representation} The algebraic engineering requires the introduction of two types of representations for the asymptotic algebra $\CU^a$. Following the terminology of the 5D construction \cite{AFS,Bourgine2017b}, we will call them \textit{vertical} and \textit{horizontal}. The vertical representation is an infinite dimensional highest $\ell$-weight representation called the \textit{prefundamental} representation \cite{Hernandez2012}. In this representation, that we denote $\rho_\nu$, the Drinfeld currents act on the module $\CL_\nu$ spanned by vectors $\dket{k}$ labeled by positive integers $k$,
\begin{align}\label{prefund}
\begin{split}
&\rho_\nu(X^+(z))\dket{k}=\dfrac1{q-q^{-1}}\d(\nu q^{2k}/z)\dket{k+1},\\
&\rho_\nu(X^-(z))\dket{k}=\dfrac{q^{-1}}{q-q^{-1}}\d(\nu q^{2(k-1)}/z)(1-q^{2k})\dket{k-1},\\
&\rho_\nu(\Psi^\pm(z))\dket{k}=q^{2k}\left[\dfrac{z(z-\nu q^{-2})}{(z-\nu q^{2k})(z-\nu q^{2(k-1)})}\right]_\pm\dket{k}.
\end{split}
\end{align}
The subscripts on the square brackets $[\cdots]_\pm$ refer to the expansions in powers of $z^{\mp1}$. This representation has weight $\nu\in\mC^\times$ and level $c=0$.\footnote{It is sometimes denoted $L_{1,\nu}^-$ which refers to the action of the Cartan subalgebra on the vacuum state
\begin{equation}
\rho_\nu(\Psi^\pm(u^{-1}))\dket{0}=\left[\dfrac1{1-\nu u}\right]_\pm\dket{0}.
\end{equation}}
Expanding the Cartan currents, we deduce that the action of $\Psi_0^-$ is vanishing on every states, and the prefundamental representation cannot be defined for the algebra $\CU$. It is, however, a well-defined representation for the shifted algebra $\CU^{(0,-1)}$ since $\Psi_0^+\Psi_{-1}^-=\Psi_{-1}^-\Psi_0^+=(-\nu)^{-1}\in\mC^\times$. In fact, the prefundamental representation is the formal limit $N\to\infty$ of the $(N+1)$-dimensional Kirillov-Reshetikhin modules of $\CU$ (a.k.a. the representations of spin $N/2$ in physics). In fact, these modules can be recovered as a submodule for the shifted prefundamental representation $\iota_{P_N}\rho_{\nu}$ where the polynomial $P_N(z)=q^{-N}(1-\nu q^{2N}/z)$ acts as a sort of projector inside $X^+(z)$ whose action now vanishes on the state $\dket{N}$.

We introduce the dual states $\dbra{k}$ by imposing the contragredient action $X^\pm(z)^\dagger=-X^\mp(z)$ and $\Psi^\pm(z)^\dagger=\Psi^\pm(z)$. It fixes the scalar product to
\begin{equation}\label{norm_ketN}
\dbra{k}\!\!\dket{k'}=n_k(\nu)^{-1}\d_{k,k'},\quad\text{with}\quad n_k(\nu)=(-1)^kq^{k}(q^2;q^2)_k^{-1}.
\end{equation} 

\paragraph{Vortices and Vermas}\footnote{This paragraph develops a side remark and can be skipped on first reading.} It is instructive compare the prefundamental module $\CL_\nu$ with the Verma modules of vortices appearing in the context of 3D $\CN=4$ gauge theories \cite{Bullimore2016}. When these theories are considered on the omega-background $\mR_\e^2\times \mR$, the chiral ring $\mC[\CM_C]$ of their Coulomb branch becomes a non-commutative algebra $\mC_\e[\CM_C]$ \cite{Bullimore2015,Braverman2016}. The equivariant cohomology of the Coulomb branch defines a Hilbert space that is a Verma module for the quantized chiral ring $\mC_\e[\CM_C]$. As we shall see, the prefundamental module coincides with the Verma module for an abelian 3D $\CN=4$ gauge theory with $N$ hypermultiplets, and the shifted representation $\iota_P\rho_\nu$ with the polynomial $P(z)=\prod_{a>2}(1-z/(q^{-2}\nu\fund_a))$ generates a q-deformed (or K-theoretic) version of the quantum chiral ring action. The connection between the 3D $\CN=4$ theories on $\mR_\e^2\times \mR_t$ and the 3D $\CN=2$ theories on $\mR_\e^2\times S_R^1$ discussed in this paper involves 2D $\CN=(2,2)$ theories on $\mR_\e^2$. The latter define half-BPS Neumann boundary conditions at fixed time $t$ for the 3D $\CN=4$ theories, they correspond to certain Whittaker vectors in the Verma module. Moreover, they are simply the KK reduction $R\to0$ of the 3D $\CN=2$ theories on $\mR_\e^2\times S_R^1$ (hence the q-deformation of the quantum chiral ring action). Thus, the action $\iota_P\rho_\nu$ of the shifted quantum affine $\sl(2)$ algebra on the equivariant cohomology of the Coulomb branch $\CM_C$ can be interpreted as the q-deformed version of the ``finite AGT-correspondence'' involving the shifted Yangian of $\sl(2)$ \cite{Braverman2010}.

The action of $\iota_P\rho_\nu$ on $\CL_\nu$ takes the following form after a simple rescaling of the states $\dket{k}\to (q^2;q^2)_k^{-1}\dket{k}$,
\begin{align}
\begin{split}
&\iota_P\rho_\nu(X^+(z))\dket{k}=\dfrac1{q-q^{-1}}\d(\nu q^{2k}/z)P(\nu q^{2k})(1-q^{2k+2})\dket{k+1},\\
&\iota_P\rho_\nu(X^-(z))\dket{k}=\dfrac{q^{-1}}{q-q^{-1}}\d(\nu q^{2(k-1)}/z)\dket{k-1},\\
&\iota_P\rho_\nu(\Psi^\pm(z))\dket{k}=q^{2k}\left[P(z)\dfrac{z(z-\nu q^{-2})}{(z-\nu q^{2k})(z-\nu q^{2(k-1)})}\right]_\pm\dket{k}.
\end{split}
\end{align}
This action can also be expressed using the q-oscillators of Bazhanov-Lukyanov-Zamolodchikov \cite{BLZ3,Boos2015}, i.e. using the operators $b,b^\dagger$ and $K$ satisfying 
\begin{equation}
q^{\a K}b^\dagger=q^\a b^\dagger q^{\a K},\quad q^{\a K}b=q^{-\a} b q^{\a K},\quad b^\dagger b=[K]_q,\quad bb^\dagger=[K+1]_q,
\end{equation} 
with the $q$-numbers $[x]_q=(q^x-q^{-x})/(q-q^{-1})$. These operators act on the states of the module $\CL_\nu$ as $q^{K}\dket{k}=q^k\dket{k}$, $b^\dagger\dket{k}=[k+1]_q\dket{k+1}$, and $b\dket{k}=\dket{k-1}$. They can be used to rewrite the action of the modes of the Drinfeld currents (omitting the representation $\iota_P\rho_\nu$)\footnote{As a result, since $\nu_1\fund=\nu$,
\begin{align}
\begin{split}
&X_n^+X_m^-=\dfrac{q^{-1}(\nu q^{2K-2})^{n+m}}{(q-q^{-1})^2}P(\nu q^{2K-2})(1-q^{2K})=\dfrac{q^{-1}(\nu q^{2K-2})^{n+m}}{(q-q^{-1})^2}\prod_{a=1}^{N\fund}(1-\nu q^{2K}/\nu_a\fund),\\
&X_n^-X_m^+=\dfrac{q^{-1}(\nu q^{2K})^{n+m}}{(q-q^{-1})^2}P(\nu q^{2K})(1-q^{2K+2})=\dfrac{q^{-1}(\nu q^{2K})^{n+m}}{(q-q^{-1})^2}\prod_{a=1}^{N\fund}(1-\nu q^{2K+2}/\nu_a\fund).
\end{split}
\end{align}}
\begin{align}
\begin{split}
&X_n^+=-q\nu^n b^\dagger q^{(2n+1)K}P(\nu q^{2K}),\quad X_n^-=\dfrac{q^{-1}}{q-q^{-1}}\nu^nq^{2n\hat K}b,\\
\end{split}
\end{align}
while the action of the Cartan modes $\Psi_{\pm n}^\pm$ only involves the operator $K$. It leads to identify $K$ with the action of the adjoint complex scalar field $\vphi$ of the $\CN=4$ vector multiplet (up to a shift involving $m_1=\log\nu$). The zero modes $X_0^+\sim b^\dagger P(\nu q^{2K})$ and $X_0^-\sim b$ are the q-deformed versions of the monopole operators $v_\pm$, while the other modes correspond to monopole operators dressed by (gauge-invariant) polynomials of $\vphi$. In this simple case, the corresponding 2D $\CN=(2,2)$ theory has a $U(1)$ gauge group and $N$ fundamental chiral multiplets. It is associated to the Lagrangian splitting $(+,\cdots,+)$ of the 3D $\CN=4$ hypermultiplets. Other choices of splitting are obtained using both shift morphisms $\iota_{P_+}\iota_{P_-}^\ast\rho_\nu$.

We will not discuss the non-abelian case here, but it can be treated using to coproduct \ref{def_coprod}. In fact, its action is reminiscent of the abelianization map introduced in \cite{Bullimore2015}, and it would be interesting to develop further this analogy.

\paragraph{Horizontal representation} To complete the formulation of the algebraic engineering, we need to introduce a second representation of $\CU^a$ that we call ``horizontal''. In this representation, denoted $\rho_u^{(n)}$, the currents are expressed as vertex operators acting on the Fock space $\CF_u$ of a 2D free boson. As usual, this Fock space is built upon the vacuum state $\ket{\vac}$ by the action of the negative modes $J_{-k}$ of the Heisenberg algebra $[J_k,J_l]=k\d_{k+l}$, $[J_0,Q]=1$, with the vacuum $\ket{\vac}$ being anihilated by the modes $J_{k>0}$. As the notation suggests, this representation depends on a weight $u\in\mC^\times$ and an integer $n$.  The weight $u$ can be identified with the eigenvalue of the zero mode of the free boson on the vacuum state.\footnote{Strictly speaking, we should replace $u$ and $u\p_u$ with the operators $\hat u$, $p_u$ with $[p_u,\hat u]=\hat u$ that act on the vacuum state as $\bra{\vac}\hat u=u\bra{\vac}$ ($\hat u$ is normal-ordered on the left). These operators correspond to the zero-modes of the free boson $\phi(z)=p_u-\log \hat u\log z-\sum_k\frac1k z^{-k}J_k$. We slightly abuse the notation here to shorten formulas.}

The expression of the currents read
\begin{align}
\begin{split}\label{def_rho_n}
&\rho_{u}^{(n)}(X^+(z))=-\dfrac{uz^{-n}}{q-q^{-1}}\chi^-(q^{2}z)^{-1}\chi^+(z)q^{-2u\p_u},\quad \rho_{u}^{(n)}(X^-(z))=\dfrac{(qu)^{-1}z^{n}}{q-q^{-1}}\chi^-(z)^{-1}\chi^+(z)^{-1}\\
&\rho_{u}^{(n)}(\Psi^+(z))=0,\quad \rho_{u}^{(n)}(\Psi^-(z))=\chi^-(z)^{-1}\chi^-(q^{2}z)^{-1}q^{-2u\p_u},
\end{split}
\end{align}
where we used a shortcut notation for the vertex operators
\begin{align}\label{def_chi}
\begin{split}
\chi^+(z)=e^{-\sum_{k>0}\frac{z^{-k}}{k}J_k},\quad \chi^-(z)=e^{\sum_{k>0}\frac{z^k}{k}(1-q^{-2k})J_{-k}}.
\end{split}
\end{align}
It is a representation of level zero, but can be extended to any level $c$ by a twist. We will discuss further the origin of this representation and the twist in the companion paper \cite{Bourgine2021c}. In fact, the dependence in $n$ is slightly redundant since $\rho_{u}^{(n+p)}=\iota_{z^{-p}}\iota_{z^p}^\ast\rho_u^{(n)}$ but we keep it explicit here as this parameter will be used to build the Chern-Simons level of the gauge theories.

Obviously, the assignment $\rho_{u}^{(n)}(\Psi^+(z))=0$ is not compatible with the definition of the original algebra $\CU$, but it is allowed for the asymptotic algebra in which the condition $\Psi_0^+\Psi_0^-=1$ is relaxed. Alternatively, we can see the relations \ref{def_rho_n} as defining a representation for the shifted algebra $\CU^{(-\infty,0)}$ in which all the modes $\Psi_k^+$ disappear as a consequence of the infinite shift. It is noted that a similar representation of either $\CU^a$ or $\CU^{(0,-\infty)}$ on $\CF_u$, with this time $\brho_{u}^{(n)}(\Psi^+(z))=0$, can also be introduced but it will not be used here.

Attentive readers might have noticed that the expression of $X^-(z)$ in \ref{def_rho_n} coincides with Jing's ``t-fermion'' introduced in the context of Hall-Littlewood polynomials \cite{Jing1991}. This is no coincidence given the fact that the Fock representation of the quantum toroidal $\gl(1)$ algebra has a deep connection with Macdonald polynomials \cite{Feigin2009a,Zenkevich2016,Bourgine2018a}, and that the representation discussed here appears in the limit $q_\text{Mac.}=q_1^{-1}\to0$ that sends Macdonald to Hall-Littlewood polynomials. Once again, this will be discussed further in \cite{Bourgine2021c}.

\subsection{Intertwining operators}\label{sec_def_P_nu}
In this subsection, we introduce the operators $\Phi$ and $\Phi^\ast$ that intertwine between the action of the asymptotic quantum affine $\sl(2)$ algebra on the Fock module $\CF_{u'}$, and on the tensor product $\CL_\nu\otimes\CF_u$ of the Fock and prefundamental modules. We start with the operator $\Phi[u,\nu,n]:\CL_\nu\otimes\CF_u\to\CF_{(-\nu)^{1/2}u}$ defined as a sum of vertex operators,
\begin{align}\label{def_Phi}
\begin{split}
&\Phi[u,\nu,n]=\sum_{k=0}^\infty n_k\ \dbra{k}\otimes\Phi_k[u,\nu,n],\\
&\Phi_k[u,\nu,n]=\t_k[\nu,n] u^k e^{-\sum_{l>0}\frac1l\nu^l q^{2kl}J_{-l}}e^{-\sum_{l>0}\frac1l\nu^{-l}\frac{1-q^{-2kl}}{1-q^{-2l}}J_l}q^{-2ku\p_u},
\end{split}
\end{align}
with $\t_k[\nu,n]=(-)^k\nu^{-nk}q^{-(n+1)k(k-1)}$ and $n_k(\nu)=\dbra{k}\!\!\dket{k}^{-1}$ given in the r.h.s. of \ref{norm_ketN}. For simplicity, we will omit to indicate the dependence in $u,\nu,n$ when it is not strictly necessary. A short calculation shows that the operator $\Phi$ obeys the intertwining relation
\begin{equation}\label{intw_Phi}
\rho_{u}^{(n)}(e)\Phi =\Phi\ \left(\rho_\nu\otimes \iota_{P_\nu}^\ast\rho_{u}^{(n)}\ \D(e)\right),\quad \forall e\in\CU^a,
\end{equation} 
involving a shift introduced by the morphism $\iota_{P_\nu}^\ast$ with the polynomial $P_\nu(z)=1-\nu/z$. In the language of shifted algebras, $\Phi$ intertwines between the action of $\CU^{(0,-1)}\otimes\CU^{(-\infty,1)}$ and $\CU^{(-\infty,0)}$. Thus, the presence of the morphism $\iota_{P_\nu}^\ast$ is necessary in order to compensate for the fact that the prefundamental action actually represents the shifted algebra $\CU^{(0,-1)}$. The physical interpretation for this shift lies in the presence of the fundamental chiral multiplet of mass $\nu_i$ that enters in the vector contribution $\Zv(\bk,\bnu)$. In string theory, it follows from the semi-infinite D5-branes on the right of figure \ref{figHiggs} that ``dress'' the two NS5-branes nearby.

The operator $\Phi^\ast[u,\nu,n]:\CF_{-u\nu}\to\CL_\nu\otimes\CF_u$ is constructed in the same way, i.e. as an expansion on the vertical components of vertex operators
\begin{align}\label{def_Phis}
\begin{split}
&\Phi^\ast[u,\nu,n]=\sum_{k=0}^\infty n_k \dket{k}\otimes \Phi_k^\ast[u,\nu,n],\\
&\text{with}\quad \Phi_k^\ast[u,\nu,n]=\t_k^\ast[\nu,n]u^{-k}e^{\sum_{l>0}\frac{\nu^l}{l}q^{-2l}(1-q^{2kl})J_{-l}}e^{-\sum_{l>0}\frac{\nu^{-l}}{l}\frac{q^{-2kl}}{1-q^{-2l}}J_l},
\end{split}
\end{align}
and $\t_k^\ast[\nu,n]=\nu^{nk}q^{-k+(n+1)k(k-1)}$. The intertwining relations involve a new morphism $\iota_{P_\nu^\ast}$ based on a different polynomial, namely $P_\nu^\ast(z)=z(z-\nu q^{-2})q^2\nu^{-2}$,
\begin{equation}\label{intw_Phis}
\left(\rho_{\nu}\otimes\rho_{u}^{(n)}\ \D'(e)\right)\Phi^\ast=\Phi^\ast\ \iota_{P_\nu^\ast}\rho_{-u\nu}^{(n+1)}(e),\quad e\in\CU^a.
\end{equation}
This relation states that $\Phi^\ast$ intertwines between the action of $\CU^{(0,-1)}\otimes\CU^{(-\infty,0)}$ and $\CU^{(-\infty,-1)}$.

\begin{figure}
\begin{center}
\begin{tabular}{|c|c|c|}
\hline
Representations & Algebras & Intertwiner\\
\hline
$\iota_P \rho_\nu\otimes \iota_{P_\nu}^\ast\rho_u^{(n)}\to\iota_P\rho_u^{(n)}$ & $\CU^{(0,-1)+\bmu_P}\otimes\CU^{(-\infty,1)}\to\CU^{(-\infty,0)+\bmu_P}$ & $\Phi$\\
$\iota_P^\ast \rho_\nu\otimes \iota_{P_\nu}^\ast\rho_u^{(n)}\to\iota_P\rho_u^{(n)}$ & $\CU^{(0,-1)+\bmu_P}\otimes\CU^{(-\infty,1)}\to\CU^{(-\infty,0)+\bmu_P}$ & $\Phi^P$\\
$\rho_\nu\otimes\iota_{P_\nu P}^\ast\rho_u^{(n)}\to\iota_P^\ast\rho_u^{(n)}$ & $\CU^{(0,-1)}\otimes\CU^{(-\infty,1)+\bmu_P}\to\CU^{(-\infty,0)+\bmu_P}$ & $\Phi$\\
$\rho_\nu\otimes\iota_P\iota_{P_\nu}^\ast\rho_u^{(n)}\to\iota_P\rho_u^{(n)}$ & $\CU^{(0,-1)}\otimes\CU^{(-\infty,1)+\bmu_P}\to\CU^{(-\infty,0)+\bmu_P}$ & $\Phi^P$\\
$\iota_P^\ast\iota_{P_\nu^\ast}\rho_{-u\nu}^{(n+1)}\to\iota_P\rho_\nu\otimes\rho_u^{(n)}$ & $\CU^{(-\infty,-1)+\bmu_P}\to\CU^{(0,-1)+\bmu_P}\otimes\CU^{(-\infty,0)}$ & $\Phi^\ast$\\
$\iota_P^\ast\iota_{P_\nu^\ast}\rho_{-u\nu}^{(n+1)}\to\iota_P^\ast\rho_\nu\otimes\rho_u^{(n)}$ & $\CU^{(-\infty,-1)+\bmu_P}\to\CU^{(0,-1)+\bmu_P}\otimes\CU^{(-\infty,0)}$ & $\Phi^{P\ast}$\\
$\iota_{PP_\nu^\ast}\rho_{-u\nu}^{(n+1)}\to\rho_\nu\otimes\iota_P\rho_u^{(n)}$ & $\CU^{(-\infty,-1)+\bmu_P}\to\CU^{(0,-1)}\otimes\CU^{(-\infty,0)+\bmu_P}$ & $\Phi^\ast$\\
$\iota_P^\ast\iota_{P_\nu^\ast}\rho_{-u\nu}^{(n+1)}\to\rho_\nu\otimes\iota_P^\ast\rho_u^{(n)}$ & $\CU^{(-\infty,-1)+\bmu_P}\to\CU^{(0,-1)}\otimes\CU^{(-\infty,0)+\bmu_P}$ & $\Phi^{P\ast}$\\
\hline
\end{tabular}
\end{center}
\caption{Summary of the various intertwining relations for $\CU^\bmu$}
\label{table3}
\end{figure}

\paragraph{Shifted intertwiners} The horizontal ``gluing'' of intertwiners involve a product of operators. As a result, the shifts by the automorphisms $\iota_P$, $\iota_P^\ast$ will propagate along the NS5-branes. Thus, we need to examine the intertwining relations \ref{intw_Phi} and \ref{intw_Phis} when $\rho_{u}^{(n)}$ is shifted by an extra polynomial $P(z)$, using either $\iota_P$ or $\iota_P^\ast$, on both sides of the relation. Tracking the factors $P(z)$ inside these intertwining equations, we can show that in some cases the relations are satisfied by the original intertwiners $\Phi$ and $\Phi^\ast$, and in some others they have to be replaced by $P(z)$-dependent intertwiners $\Phi^P$ and $\Phi^{P\ast}$. The results of this analysis are summarized on the table of figure \ref{table3}. The modified intertwiners have the same operator part, but their vertical components contain an extra factor,
\begin{align}
\begin{split}
&\t_k^P[\nu,n]=\prod_{j=1}^k P(\nu q^{2j-2})\ \t_k[\nu,n],\\
&\t_k^{P\ast}[\nu,n]=\prod_{j=1}^k P(\nu q^{2j-2})\ \t_k^{\ast}[\nu,n].
\end{split}
\end{align}
It is worth noting that, taking the polynomial $P(z)=\prod_{a=1}^d(1-z/\mu_a)$, we have
\begin{equation}\label{P_Zaf}
\Zf(k,\nu,\bmu)=\prod_{j=1}^k P(\nu q^{2j-2})^{-1},\quad\Zaf(k,\nu,\bmu)=\prod_{j=1}^k P(\nu q^{2j-2}),
\end{equation} 
and the corresponding intertwiners can be used to introduce the contribution of chiral matter multiplets. The use of shifted intertwiners to render fundamental/antifundamental matter fields will be discuss with more details in \cite{Bourgine2021c}.

\subsection{Algebraic engineering}
In order to construct the observables of the 3D $\CN=2$ gauge theories, we need to \textit{glue} the intertwiners, which means take their product along common modules in the representations network. Along a vertical leg carrying a prefundamental representation, the gluing is realized by the scalar product $\dbra{k}\!\!\dket{k'}=n_k(\nu)^{-1}\d_{k,k'}$, whereas it is realized as a product of operators in horizontal modules. The corresponding vacuum expectation value is computed using the normal-ordering relations\footnote{In order to compute the vacuum expectation value, we need to adjust the charge on the left such that $\la e^{\b Q}\cdots q^{\a J_0}\ra=\bra{\b}e^{\b Q}\cdots q^{\a J_0}\ket{\vac}$, with $\bra{\b}=\bra{\vac}e^{-\b Q}$.}
\begin{align}\label{NO_3D}
\begin{split}
&\Phi_{k_1}[u_1,\nu_1,n_1]\Phi_{k_2}[u_2,\nu_2,n_2]=q^{-2k_1k_2}N_{k_2,k_1}(\nu_2/\nu_1)^{-1}:\Phi_{k_1}[u_1,\nu_1,n_1]\Phi_{k_2}[u_2,\nu_2,n_2]:,\\
&\Phi_{k_1}[u_1,\nu_1,n_1]\Phi_{k_2}^\ast[u_2,\nu_2,n_2]=\dfrac{N_{k_1,k_2}(\nu_1/\nu_2)}{N_{0,k_2}(\nu_1/\nu_2)}:\Phi_{k_1}[u_1,\nu_1,n_1]\Phi_{k_2}^\ast[u_2,\nu_2,n_2]:,\\
&\Phi_{k_1}^\ast[u_1,\nu_1,n_1]\Phi_{k_2}[u_2,\nu_2,n_2]=(q^2\nu_2/\nu_1;q^2)_\infty\dfrac{N_{k_2,k_1}(\nu_2/\nu_1)}{(q^2\nu_2/\nu_1;q^2)_{k_2}}:\Phi_{k_1}^\ast[u_1,\nu_1,n_1]\Phi_{k_2}[u_2,\nu_2,n_2]:,\\
&\Phi_{k_1}^\ast[u_1,\nu_1,n_1]\Phi_{k_2}^\ast[u_2,\nu_2,n_2]=(-\nu_1)^{k_2}\nu_2^{-k_2}q^{2k_1k_2-k_2(k_2-1)}N_{k_1,k_2}(\nu_1/\nu_2)^{-1}:\Phi_{k_1}^\ast[u_1,\nu_1,n_1]\Phi_{k_2}^\ast[u_2,\nu_2,n_2]:.
\end{split}
\end{align}
In this way, the representations network defines an operator $T$ that intertwines between the representations associated to the external legs. As we shall see, its vacuum expectation value reproduces the vortex partition function of a gauge theory.

The operator $T_{U(1)}$ describing the $U(1)$ gauge theory with a single fundamental chiral multiplet is the easiest to compute as it involves only a single vertical gluing. Taking the vev in the module associated to the external edges, we find
\begin{equation}
T_\text{U(1)}=\sum_{k=0}^\infty n_k(\nu)\ \Phi_k^\ast[u^\ast,\nu,n^\ast]\otimes\Phi_k[u,\nu,n]\implies \la T_\text{U(1)}\ra=\sum_{k=0}^\infty \left(\dfrac{u}{u^\ast}\right)^k \dfrac{(\nu^k q^{k(k-1)})^{n^\ast-n}}{(q^2;q^2)_k}.
\end{equation} 
To perform the identification, we rewrite the summands using $N_{k,k}(1)=(q^2;q^2)_k$. As a result, we recover indeed the vortex partition function $Z_V[U(1),1,0]$ upon identification of the exponentiated FI parameter $\qf=u/u^\ast$, and the Chern-Simons level $\k=n^\ast-n$.

\begin{figure}
\begin{center}
\begin{tikzpicture}[scale=1.2]
\draw[postaction={on each segment={mid arrow=black}}] (2.7,-0.7) -- (2,0) -- (0,0) -- (-0.7,-0.7);
\draw[postaction={on each segment={mid arrow=black}}] (3,2) -- (2,2) -- (0,2) -- (-1,2);
\draw[postaction={on each segment={mid arrow=black}}] (0,0) -- (0,2);
\draw[postaction={on each segment={mid arrow=black}}] (2,0) -- (2,2);
\node[left,scale=0.7] at (0,0) {$\Phi_1^\ast$};
\node[above,scale=0.7] at (0,2) {$\Phi_1$};
\node[above,scale=0.7] at (2,2) {$\Phi_2$};
\node[right,scale=0.7] at (2,0) {$\Phi_2^\ast$};
\node[below,scale=0.7] at (-0.7,-0.7) {$\rho_{u_1^\ast}^{(n_1^\ast)}$};
\node[below,scale=0.7] at (2.7,-0.7) {$\iota_{P_{\nu_1}^\ast P_{\nu_2}^\ast}\rho_{(\nu_1\nu_2)^{1/2}u_1^\ast}^{(n_1^\ast+2)}$};
\node[left,scale=0.7] at (-1,2) {$\rho_{(\nu_1\nu_2)^{1/2}u_2}^{(n_2)}$};
\node[right,scale=0.7] at (3,2) {$\iota_{P_{\nu_1}P_{\nu_2}}^\ast\rho_{u_2}^{(n_2)}$};
\node[above,scale=0.7] at (1,2) {$\iota_{P_{\nu_1}}^\ast\rho_{(-\nu_2)^{1/2}u_2}^{(n_2)}$};
\node[below,scale=0.7] at (1,0) {$\iota_{P_{\nu_1}^\ast}\rho_{(-\nu_1)^{1/2}u_1^\ast}^{(n_1^\ast+1)}$};
\node[left,scale=0.7] at (0,1) {$\rho_{\nu_1}$};
\node[right,scale=0.7] at (2,1) {$\rho_{\nu_2}$};
\end{tikzpicture}
\hspace{10mm}
\begin{tikzpicture}[scale=1.2]
\draw[postaction={on each segment={mid arrow=black}}] (2.7,.3) -- (2,1) -- (2,3) -- (2.7,3.7);
\draw[postaction={on each segment={mid arrow=black}}] (0,0) -- (0,1) -- (0,3) -- (0,4);
\draw[postaction={on each segment={mid arrow=black}}] (2,1) -- (0,1);
\draw[postaction={on each segment={mid arrow=black}}] (2,3) -- (0,3);
\node[left,scale=0.7] at (0,3) {$\Phi_1$};
\node[left,scale=0.7] at (0,1) {$\Phi_2$};
\node[right,scale=0.7] at (2,3) {$\Phi_1^\ast$};
\node[right,scale=0.7] at (2,1) {$\Phi_2^\ast$};
\node[above right,scale=0.7] at (2.7,3.7) {$\rho_{u_1^\ast}^{(n_1^\ast)}$};
\node[below right,scale=0.7] at (2.7,.3) {$\iota_{P_{\nu_1}^\ast P_{\nu_2}^\ast}\rho_{(\nu_1\nu_2)^{1/2}u_1^\ast}^{(n_1^\ast+2)}$};
\node[above,scale=0.7] at (0,4) {$\rho_{u_2}^{(n_2)}$};
\node[below,scale=0.7] at (0,0) {$\iota_{P_{\nu_1}P_{\nu_2}}^\ast\rho_{(\nu_1\nu_2)^{1/2}u_2}^{(n_2)}$};
\node[left,scale=0.7] at (0,2) {$\iota_{P_{\nu_1}}^\ast\rho_{(-\nu_2)^{1/2}u_2}^{(n_2)}$};
\node[right,scale=0.7] at (2,2) {$\iota_{P_{\nu_1}^\ast}\rho_{(-\nu_1)^{1/2}u_1^\ast}^{(n_1^\ast+1)}$};
\node[above,scale=0.7] at (1,3) {$\rho_{\nu_1}$};
\node[below,scale=0.7] at (1,1) {$\rho_{\nu_2}$};
\end{tikzpicture}
\end{center}
\caption{Network of representations for the $U(2)$ partition function, and a rotated version that can be superimposed on the branes web.}
\label{fig7}
\end{figure}
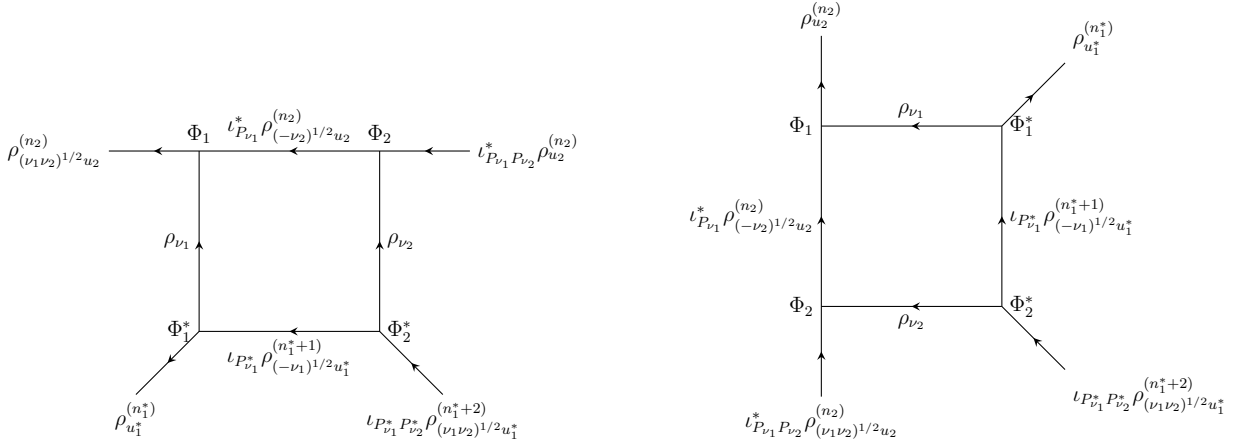

The $U(2)$ partition function is a little harder to derive, it is obtained after two vertical gluings and two horizontal ones, according to the representations network indicated on figure \ref{fig7}. The corresponding operator $T_\text{U(2)}$ reads
\begin{align}
\begin{split}\label{ZV_U2}
&T_\text{U(2)}=\sum_{k_1,k_2=0}^\infty n_{k_1}(\nu_1)n_{k_2}(\nu_2)\ \Phi^\ast_{k_1}[u^\ast_1,\nu_1,n^\ast_1]\Phi^\ast_{k_2}[u^\ast_2,\nu_2,n^\ast_2]\otimes\Phi_{k_1}[u_1,\nu_1,n_1]\Phi_{k_2}[u_2,\nu_2,n_2],\\
&\implies\la T_\text{U(2)}\ra=\sum_{k_1,k_2=0}^\infty \qf_1^{k_1}\qf_2^{k_2}\dfrac{\ZCS(k_1,\nu_1,\k_1)\ZCS(k_2,\nu_2,\k_2)}{\prod_{i,i'=1}^2 N_{k_i,k_{i'}}(\nu_i/\nu_{i'})}.
\end{split}
\end{align}
We recognize here $Z_V[U(2),2,0]$ after the identification of the parameters $\qf_1u_1/u_1^\ast$, $\qf_2=-\nu_1u_2/u_2^\ast$, $\k_1=n_1^\ast-n_1$ and $\k_2=n_2^\ast-n_2-1$. The compatibility between representations imposes the constraints $u_2=u_1$, $u_2^\ast=-u_1^\ast\nu_1$ and $n_2=n_1$, $n_2^\ast=n_1^\ast+1$, so that $\qf_1=\qf_2$ and $\k_1=\k_2$.

Upon a rotation (see figure \ref{fig7} (right)), the network of representations can be superimposed to the brane systems of figure \ref{figHiggs} at $\a=0$, thereby assigning the horizontal modules to NS5-branes (possibly dressed by D5-branes), and the prefundamental modules to D3-branes. Just like in the original $(p,q)$-branes web, the weights $\log u$ and $\log u^\ast$ encode the position of the NS5-branes along the $x^6$ direction, and their difference corresponds to the (renormalized) FI parameter. The position of the D3 branes along the $x^5$ direction, which is the same as the D5-branes on the left, is identified with the logarithm of the prefundamental weights $\nu_a$. The intertwiner $\Phi$ describes a D3-brane ending on a NS5-brane, it leaves the parameter $n$ unchanged as the D3-brane carries no $(p,q)$-charge. From the point of view of the 6D $\CN=(1,1)$ theory on the NS5-branes, this D3-branes is a magnetic monopole. On the other hand, the intertwiner $\Phi^\ast$ describes the combination of D3+D5 branes ending on a dressed NS5-branes (i.e. a $(1,n)$-brane). Due to the charge of the D5-brane, the parameter $n$ is shifted to $n+1$. Finally, as in the $(p,q)$-brane web, the difference of angle between dressed NS5-branes (after projection in the (56)-plane) gives the Chern-Simons level.

This construction trivially generalizes to the case of $U(N)$ with $N_f=N$ fundamental chiral multiplets. Moreover, fundamental/antifundamental matter chiral multiplets can be introduced by shifting some of the horizontal representations, using the morphisms $\iota_P$ or $\iota_P^\ast$, so that the intertwiners are replaced by their shifted version $\Phi^P$ and $\Phi^{P\ast}$ (according to the table of figure \ref{table3}), the matter contribution following from the equalities \ref{P_Zaf}. Finally, linear quiver gauge theories with $\times_a U(N_a)$ gauge groups can also be constructed in a similar way. We will perform these constructions explicitly in the companion paper \cite{Bourgine2021c}.

\paragraph{qq-character} We would like to conclude this section by presenting the derivation of the fundamental qq-character which proceeds from the same method as the one exposed in \cite{Bourgine2017b} for 5D $\CN=1$ gauge theories. There, the fundamental qq-character is obtained  algebraically by the insertion of the operators $\D X^\pm(z)$ on either the left or the right side of the diagram (since these operators commute with the operator $T$ by construction),
\begin{equation}\label{char_U2}
\la\left(\rho_{u_1^\ast}^{(n_1^\ast)}\otimes\rho_{u_1}^{(n_1)}\ \D(X^\pm(z)\right)T_{U(2)}\ra=\la T_{U(2)}\left(\iota_{P_{\nu_1}^\ast P_{\nu_2}^\ast}\rho_{u_1^\ast\nu_1\nu_2}^{(n_1^\ast+2)}\otimes\iota_{P_{\nu_1}P_{\nu_2}}^\ast\rho_{u_1}^{(n_1)}\ \D(X^\pm(z)\right)\ra.
\end{equation}
Due to this commutation property, the qq-character is a polynomial in $z$. In the 3D case studied here, the term $X^-(z)\otimes\Psi^+(z)$ of $\D(X^-(z))$ is projected out by the representations, and we find the simple result
\begin{equation}
\dfrac{(qu_1)^{-1} z^{n_1-2}}{q-q^{-1}}\sum_{k_1,k_2=0}^\infty \qf^{|\bk|}\Zv(\bk,\bnu)\ZCS(\bk,\bnu,\k)\ (z-\nu_1q^{2k_1})(z-\nu_2q^{2k_2})
\end{equation} 
which is trivially a polynomial of degree two in $z$ up to an overall power $z^{n_1-2}$. By contrast, the insertion of the operator $\D(X^+(z))$ remains non-trivial since the term $\Psi^-(z)\otimes X^+(z)$ is not projected out,
\begin{align}\label{qq_char}
\begin{split}
&-\dfrac{q^{-2} u_1^\ast  z^{-n_1^\ast}}{q-q^{-1}}\sum_{k_1,k_2=0}^\infty\qf^{|\bk|}\Zv(\bk,\bnu)\ZCS(\bk,\bnu,\k)\ \left[Y_{\bk}^\ast(q^2z)+\qf\dfrac{z^{\k}}{Y_{\bk}(z)}\right].
\end{split}
\end{align}
Forgetting the prefactor, this expression coincides with the limit of the fundamental vortex qq-character of the 3D $\CN=4$ $U(2)$ gauge theory with two flavors \cite{Haouzi2020}, it is a polynomial of degree $N=2$ when $\k=0$. This qq-character corresponds to the vortex partition function in the presence of a codimension two defect (wrapping $S^1$) supporting a $\CN=(2,2)$ quantum mechanics.\footnote{The existence of 2D and 3D versions of the 5D qq-characters \cite{NPS,Nekrasov_BPS1} was first conjectured in \cite{Nekrasov_BPS4,Haouzi2019}.} The functions $Y_{\bk}(z)$ and $Y_{\bk}^\ast(z)$ are the equivalent of Nekrasov's $\CY$-observables,
\begin{align}
\begin{split}\label{def_Y_obs}
Y_{\bk}(z)=\prod_{i=1}^N\left(1-\dfrac{\nu_iq^{2k_i}}{z}\right),\quad Y_{\bk}^\ast(q^2z)=\prod_{i=1}^Nq^{2k_i}\dfrac{z-\nu_iq^{-2}}{z-\nu_iq^{2k_i-2}}.
\end{split}
\end{align}
They can be obtained as a variation of the equivariant character of the single node handsaw quiver variety used in \cite{Fujitsuka2013,Yoshida2011}, and enter in the definition of its (K-theoretic) Cohomological Hall algebra \cite{Bourgine2021c}. Presumably, higher qq-characters can be derived algebraically using the method developed in \cite{Bourgine2017b} and based on the i-Weyl reflection of coproducts.

To conclude this paragraph, we would like to make a short remark regarding the connection with integrable systems. It can be shown that the fundamental vortex qq-character \ref{qq_char} provides the Baxter T-polynomial for the integrable system associated to our 3D $\CN=2$ theory in the correspondence introduced in \cite{Gadde2013}. In general, these systems are believed to be the spectral dual of the usual integrable systems of the Bethe/gauge correspondence \cite{NS2009'}. The corresponding Bethe equations can be obtained by extremizing the vortex partition function over the charges $\bk$ following the same method used for instanton sums in \cite{NPS,Bourgine2017c}.

\section{Higgsing and shifted representations}
In this section, we discuss briefly the relation between the construction method presented here and those used previously in the literature. The latter are based on the quantum toroidal $\gl(1)$ algebra with the parameters $(q_1,q_2)=(e^{R\e_1},e^{R\e_2})$ associated to the omega-deformed spacetime $\mR^2_{\e_1}\times\mR^2_{\e_2}\times S^1$. As a current algebra, the quantum toroidal $\gl(1)$ algebra reduces to the asymptotic quantum affine $\sl(2)$ algebra in the limit $q_1\to0$ (or $\e_1\to-\infty$) with $q_2=q^2$ fixed.\footnote{It must be emphasized that this is true at the level of the Drinfeld currents $x^\pm(z)$ and $\psi^\pm(z)$ but it is not true for their modes. This limit is studied for various representations in the companion paper \cite{Bourgine2021c}.} It explains the relevance of the quantum affine $\sl(2)$ algebra with quantum group parameter $q^2=e^{R\e}$ in the simpler case of the omega-background $\mR^2_{\e}\times S_R^1$. It should be noted that the limit $q_1\to\infty$ with $q_2$ fixed is equivalent to $q_1\to0$ with $q_2$ fixed thanks to the algebras morphism $\s_V$ defined in \cite{BFMZ} and its quantum affine equivalent $\s: X^\pm(z)\to X^\mp(z)$, $q^c\to q^{-c}$ considered in \cite{Hernandez2020}, both inverting the parameters $q,q_1,q_2$.

We start our discussion with the Higgsing method used in \cite{Nedelin2017,Aprile2018}, and for which the string theory description has been briefly recalled in section \ref{sec_string}. In this case, the D5-branes carry a vertical representation corresponding to the Fock representation of level $(0,1)$ built in \cite{feigin2011quantum}. It acts on a free boson Fock space with a natural basis $\dket{P_\l}$ associated to Macdonald polynomials. In the presence of an extra fundamental hypermultiplet of mass $\mu$, this action is shifted by the polynomial $P(z)=1-zq_1q_2/\mu$ using the toroidal equivalent of the morphism $\iota_P$.\footnote{Shifted quantum toroidal algebras where briefly mentioned in \cite{Finkelberg2017a}. The relation with the Higgsing of 5D $\CN=1$ theories will be exposed in \cite{Bourgine2021c} (see also \cite{Rapcak2020,Galakhov2021} in the context of affine Yangians).} As a result, the algebra now acts as follows on the Macdonald states,
\begin{align}\label{def_vert_rep}
\begin{split}
&\iota_P\rho_v^{(0,1)}(x^+(z))\dket{P_\l}=\sum_{\sAbox\in A(\l)}\delta(z/\chi_\sAbox)P(\chi_\sAbox)\res_{z=\chi_\sAbox}\dfrac1{z\CY_{\lambda}(z)}\dket{P_{\l+\sAbox}},\\
&\iota_P\rho_v^{(0,1)}(x^-(z))\dket{P_\l}=(q_1q_2)^{1/2}\sum_{\sAbox\in R(\l)}\delta(z/\chi_\sAbox)\res_{z=\chi_\sAbox}z^{-1}\CY_{\l}(q_3^{-1}z)\dket{P_{\l-\sAbox}},\\
&\iota_P\rho_v^{(0,1)}(\psi^\pm(z))\dket{P_\l}=(q_1q_2)^{1/2}\left[P(z)\dfrac{\CY_\l(q_3^{-1}z)}{\CY_\l(z)}\right]_\pm\dket{P_\l},
\end{split}
\end{align}
where the matrix elements are given by residues of the $\CY$-observable that can be found, e.g. in \cite{Bourgine2017b,Bourgine2018a}.\footnote{We recall that $A(\l)$ and $R(\l)$ denote respectively the sets of boxes that can be added or removed to the Young diagram $\l$. To each box $\Abox=(i,j)\in\l$ is associated the content $\chi_\sAbox=vq_1^{i-1}q_2^{j-1}$ depending on the weight $v$ of the representation.} When the mass takes the critical value $q_1^{-1}q_2^{-1}\mu=vq_1$, the representation becomes reducible and admit a subrepresentation acting on Young diagrams $\l=k$ built as a single column of $k$ boxes,
\begin{align}
\begin{split}
&\iota_P\rho_v^{(0,1)}(x^+(z))\dket{P_k}=(1-q_1^{-1})\delta(vq_2^k/z)q_2^k\dket{P_{k+1}},\\
&\iota_P\rho_v^{(0,1)}(x^-(z))\dket{P_k}=(q_1q_2)^{1/2}(1-q_1^{-1})\delta(vq_2^{k-1}/z)(1-q_2^{-k})\dket{P_{k-1}},\\
&\iota_P\rho_v^{(0,1)}(\psi^\pm(z))\dket{P_k}=(q_1q_2)^{1/2}\left[\dfrac{(z-vq_1q_2^k)(z-vq_1^{-1}q_2^{k-1})(z-vq_2^{-1})}{(-vq_1)(z-vq_2^k)(z-vq_2^{k-1})}\right]_\pm\dket{P_k}.
\end{split}
\end{align}
In the limit $q_1\to0$ or $\infty$, and upon identifying $q^2=q_2$ and $\nu=v$, the action of the currents $x^\pm(z)$ and $\psi^\pm(z)$ reduce to the action of $X^\pm(z)$ and $\Psi^\pm(z)$ in the prefundamental representation \ref{prefund} (up to simple rescalings). This observation elucidates the origin of the prefundamental representation in the description of vortices. Unfortunately, the limit $q_1\to\infty$ of the horizontal representation is more involved, and only a single current $x^-(z)$ survives, reproducing the vertex operator $\chi^-(z)^{-1}\chi^+(z)^{-1}$ used in \ref{def_rho_n}. It is worth noting, however, than in both constructions the module associated to the NS5-branes is a free boson Fock space, with a basis labeled by Young diagrams, and the difference comes only from which algebra is acting on it. Thus, the formalism developed here appears naturally as a consequence of the Higgsing procedure applied to the algebraic engineering of 5D $\CN=1$ gauge theories.

In the Higgsed network calculus, the intertwiners are obtained using the vector representation $\rho_v^{(0,0)}$ of the quantum toroidal $\gl(1)$ algebra instead of the Fock representation $\rho_v^{(1,0)}$ for the vertical leg \cite{Zenkevich2018}. In contrast with the latter, the vector representation is not of highest weight type, it acts on states $\dket{k}$ labeled by an integer $k\in\mZ$. The precise connection with our formalism is not fully clear at this stage. However, we can observe that the shifted vector representation $\iota_P^\ast\rho_v^{(0,0)}$ acting as
\begin{align}
\begin{split}
&\iota_P^\ast\rho_v^{(0,0)}(x^+(z))\dket{k}=(1-q_1^{-1})\d(vq_2^k/z)\dket{k+1},\\
&\iota_P^\ast\rho_v^{(0,0)}(x^-(z))\dket{k}=(1-q_1)P(vq_2^{k-1})\d(vq_2^{k-1}/z)\dket{k-1},\\
&\iota_P^\ast\rho_v^{(0,0)}(\psi^\pm(z))\dket{k}=P(z)\left[\dfrac{(z-vq_1q_2^k)(z-vq_1^{-1}q_2^{k-1})}{(z-vq_2^k)(z-vq_2^{k-1})}\right]_\pm\dket{k},
\end{split}
\end{align}
can be restricted to a highest weight representation on states $k\geq0$ if we take $P(z)=q_2^{-1}(1-q_2 z/v)$ (since then $x^-(z)\dket{0}=0$). This procedure should correspond to the choice of integration contour surrounding the poles at $z=vq_2^k$ with $k\geq0$ in \cite{Zenkevich2018}. Unsurprisingly, this representation reduces to the prefundamental representation in the limits $q_1\to0$ or $\infty$ with $q_2$ fixed. Our 3D intertwiners \ref{def_Phi} and \ref{def_Phis} are expected to be related in this way to the Higgsed intertwiners defined in \cite{Zenkevich2018}.

Alternatively, taking the representation $\iota_{Q}^\ast\rho_v^{(0,0)}$ with $Q(z)=z-vq_2^{-1}$, the current $x^+(z)$ now annihilates the state $\dket{-1}$, and we can restricted ourselves to a subrepresentation acting on the states $\dket{k}$ with $k<0$. It is tempting to associated this new representation to the other choice of contour, i.e. the contour surrounding the poles $z=vq_2^{k}$ with $k<0$. The choice of contour/representation depends on the sign of the real FI parameter $\z_\mR$. Using the map $\dket{k}\to\dket{-1-k}$, the subrepresentation $\iota_{Q}^\ast\rho_v^{(0,0)}$ can also be formulated as an action on the prefundamental module $\CL_\nu$. However, in this case the currents $x^\pm(z)$, $\psi^\pm(z)$ tend in the limit $q_1\to0$ or $\infty$ to the images $\CS^2\cdot X^\pm(z)$ and $\CS^2\cdot\Psi^\pm(z)$ of the currents acting in \ref{prefund} (and the weight is now $\nu=v^{-1}q_2^2$). The automorphism $\CS^2$ is the square of Miki's automorphism \cite{Miki2007} that (unlike $\CS$) also defines an automorphism of the quantum affine $\sl(2)$ algebra, $\CS^2\cdot X^\pm(z)=X^\mp(z^{-1})$, $\CS^2\cdot \Psi^\pm(z)=\Psi^\mp(z^{-1})$, $\CS^2\cdot q^c=q^{-c}$. It could indicate that this automorphism $\CS^2$ has a role to play in the wall-crossing phenomenon appearing when $\z_\mR$ moves from the chambers $\z_\mR>0$ to $\z_\mR<0$. We hope to be able to come back to this interesting point in a near future.

\section{Discussion}
In this letter, we have presented a formalism for the algebraic engineering of 3D $\CN=2$ gauge theories that is arguably simpler than the previous ones based on quantum toroidal algebras. We hope that it will help elucidates certain aspects of this technique, such as the role of the auxiliary Fock space supported by NS5-branes, or the relation with integrable hierarchies recently pointed out in \cite{Bourgine2021a}. The technique easily extends to type A quiver gauge theories with $U(N)$ gauge groups via the gluing of intertwiners. Other types of quivers / gauge groups might be addressed using an equivalent of the reflection states introduced in \cite{BFMZ}. Notably, our method is expected to also apply to non-Lagrangian theories possessing a brane description.

To conclude, we would like to discuss a competing algebraic technique based on W-algebras instead of quantum groups \cite{Kimura2015,Kimura2016,Kimura2017,Kimura:2019gon}. These W-algebras can be seen as a coarse-grained version of the quantum group actions as they are dual to the tensor product of representations attached to the external edges in the representations network \cite{Bourgine2018a}. The corresponding technique has produced a number of important results \cite{Kimura2016,Kimura2017,Kimura:2019gon}. It has been used recently to analyzed more general Higgsed 5D $\CN=1$ configurations with $\mu\fund_i=v_iq_1^{r_i}q_2^{s_i}$ that pertain to intersecting defects \cite{Kimura2021}. It would be interesting to revisit this analysis using the quantum group approach where this condition defines a pit of level $(r_i,s_i)$ for the shifted vertical Fock representation. For this purpose, we need to clarify the connection between the W-algebras and the quantum groups in the context of 3D gauge theories. A first step in this direction would be to identify the W-algebra arising from the tensor product of two horizontal representations $\rho_u^{(n)}$.\footnote{This W-algebra is the equivalent of the Kimura-Pestun W-algebra \cite{Kimura2015} which is a priori different from the W-algebra of the ``finite AGT correspondence'' \cite{Braverman2010} due to the loss of Miki's automorphisms \cite{Miki2007} (see e.g. \cite{Bourgine2018a,Fukuda2019}).} In this particular case, we are expecting to recover the limit of the q-Virasoro algebra studied in \cite{Ohkubo2015}. 

\section*{Acknowledgements}
The author would like to thank Sasha Garbali, David Hernandez, Saebyeok Jeong, Kimyeong Lee, Jaewon Song and Gufang Zhao for very helpful discussions. This research was partly supported by the Basic Science Research Program through the National Research Foundation of Korea (NRF) funded by the Ministry of Education through the Center for Quantum Spacetime (CQUeST) of Sogang University (NRF-2020R1A6A1A03047877). The author also gratefully acknowledge support from the Australian Research Council Centre of Excellence for Mathematical and Statistical Frontiers (ACEMS), and from the Korea Institute for Advanced Study (KIAS) for his visit of the institute.

\appendix

\bibliographystyle{../utphys}
{\footnotesize \bibliography{DIM_crystal}}
\end{document}